\documentstyle[12pt]{article}
\begin{document}
\begin{flushright}
Preprint IHEP 95-63
\end{flushright}
\begin{center}
{\bf \large Scaling Relations in Phenomenology \\
of QCD Sum Rules for Heavy Quarkonium }\\
\vspace*{0.4cm}
{V.V.Kiselev}\\
{\it Institute for High Energy Physics,\\
Protvino, Moscow Region, 142284, Russia\\
E-mail: kiselev@mx.ihep.su, vkisselev@vxcern.cern.ch\\
Fax: +7-095-230-23-37}
\end{center}
\begin{abstract}
In the framework of a specific scheme of the QCD sum rules
for $S$-wave levels of the heavy quarkonium, one derives expressions,
relating the leptonic constants, the energetic density of quarkonium states
and universal characteristics in the heavy quarkonium physics, such as
the difference between the masses of
heavy quark $Q$ and meson $(Q\bar q)$ and the number
of heavy quarkonium levels below the threshold of $(Q\bar Q) \to
(Q\bar q) + (\bar Q q)$ decay.
\end{abstract}

\section*{Introduction}

Powerful tools in studies of heavy quarkonia, the bound states of two heavy
quarks, are phenomenological potential models \cite{1,2,3} and QCD sum rules
\cite{4}. An applicability of the approaches to the systems of two heavy
quarks is caused by

1) a low value of the ratio $\Lambda_{QCD}/m_Q \ll 1$, where $m_Q$ is the
heavy quark mass and $\Lambda_{QCD}$ is the quark confinement scale,
determining the inverse distance between the quarks in the bound states, and

2) a nonrelativistic motion of the heavy quarks inside the quarkonium,
$v \to 0$.

In the QCD sum rules, the low value of ratio $\Lambda_{QCD}/m_Q$ determines
not large contribution of higher orders of the QCD perturbation theory over the
quark-gluon coupling $\alpha_S \sim 1/\ln(m_Q/\Lambda_{QCD})\ll 1$ in
the expansion of Wilson's coefficients, and it makes a suppression
of nonperturbative quark-gluon condensate contribution, having the power form,
as $O(\Lambda^4_{QCD}/m_Q^2)$ for the gluonic condensate $\langle\alpha_S \;
G^2_{\mu\nu}\rangle$ contribution into the sum rules for vector currents,
for example, at some low and moderate values of a number for the moments of
the spectral density \cite{4,v1,v11}.

As has been shown in ref.\cite{v2}, the nonperturbative effects in the heavy
quarkonium spectroscopy can not be described by a potential model, since
the correct description must take into account the retardation of
interaction with condensates. Nevertheless, the potential models are suitable
for the phenomenological studies, because they can reproduce the model
formulae or numbers for the quantities, used as input values (the level masses,
for example). Hence, the potential models can be considered as phenomenological
meaningful fittings of some experimental values, but they can not restore
a true potential, that does not exist due to the nonperturbative effects.

In the potential models, from data on the spectroscopy of the $(\bar c c)$
charmonium and the $(\bar b b)$ bottomonium one finds that the nonrelativistic
quark motion allows one to get the phenomenological potential in the range of
average distances between the heavy quarks inside the quarkonia
\begin{equation}
0.1\; fm < r < 1\;fm\;. \label{n1}
\end{equation}

Being the potential of static sources for the gluon field, this potential must
not depend on the flavours of sources. This flavour-independence is empirically
confirmed for the QCD motivated potentials \cite{1}. Such potentials,
possessing different asymptotic properties in the regions of $r \to 0$ and
$r \to \infty$, coincide with each other in the region (\ref{n1}), where
they allow approximations, having a simple scaling behaviour. These
approximations are the logarithmic \cite{2} and power \cite{3} laws
\begin{eqnarray}
V_L(r) & = & c_L + d_L \ln(\Lambda_L r)\;, \\
V_M(r) & = & c_M + d_M (\Lambda_M r)^k\;.
\end{eqnarray}
By the virial theorem
\begin{equation}
\langle T\rangle = \frac{1}{2} \langle r \frac{dV}{dr}\rangle\;,
\end{equation}
one finds
\begin{eqnarray}
\langle T_L\rangle & = & d_L/2 = const.\;, \label{99n}\\
\langle T_M\rangle & = & \frac{k}{k+2}(-c_M+E)\;, \label{10n}
\end{eqnarray}
where $E$ is the binding energy of quarks in the quarkonium.
Phenomenologically, one has $k\ll 1$, $|E|\ll |c_M|$, so that in the region
of average distances between the heavy quarks in the heavy quarkonium
(\ref{n1}), the kinetic energy of quarks practically is a constant value,
independent of the quark flavours,
\begin{equation}
\langle T_M\rangle \simeq const. \label{11n}
\end{equation}
Then from the Feynman-Hellmann theorem
\begin{equation}
\frac{dE}{d\mu} = -\;\frac{\langle T\rangle}{\mu}\;, \label{12n}
\end{equation}
where $\mu$ is the reduced mass of heavy quark system $(Q\bar Q')$,
one can get that the level difference in the system does not depend
on the reduced mass of quarks, i.e. on the quark flavours,
\begin{equation}
E(\bar n, \mu) - E(n,\mu) = E(\bar n, \mu') - E(n, \mu')\;. \label{13n}
\end{equation}
Condition (\ref{13n}) means that the energetic density of heavy quarkonium
levels does not depend on the quark flavours
\begin{equation}
\frac{dn}{dM_n} = \phi(n)\;,\label{14n}
\end{equation}
where $\phi(n)$ does not depend on $\mu$. The Bohr--Sommerfeld procedure
for the quantization of the nonrelativistic systems with the
logarithmic and power potentials gives
\begin{eqnarray}
\frac{dM_n^L}{dn} & = & \frac{2 T}{n} = \frac{1}{n}\; \frac{dM_n}{dn}(n=1)\;,
\label{r11}\\
\frac{dM_n^M}{dn} & = & V_0\; \frac{1}{n}\;
\biggl(\frac{n^2 \mu_0}{\mu}\biggr)^{k/(2+k)}\;.
\end{eqnarray}
Since $k\ll 1$, the $S$-state density in the Martin potential only
logarithmically depends on the reduced mass $\mu$. With the same accuracy,
one can state that eq.(\ref{r11}) is approximately valid in the system with
the Martin potential.

The described properties of heavy quark potential are found phenomenologically.
They cause the high accuracy of potential models for calculations of
the heavy quarkonium masses with no account of spin-dependent splittings,
$\delta m(nL) \simeq 30$ MeV. Note once more, that the scaling properties
(\ref{99n}), (\ref{11n}) and (\ref{r11}) are the leading approximations
in the fitting of the experimentally known masses of the heavy quarkonium
levels.

The accuracy for the predictions of quarkonium wave functions in the framework
of the potential models is low, for example, it is $\delta \Psi(0)/ \Psi(0)
\sim 30\div 50 \%$, since in this case the potential behaviour in the border
points ($r\to 0$ and $r\to \infty$) becomes essential. One must remember
also, that the potential, suitable for the fitting of the level masses,
can not strictly used for the potential description of the leptonic
widths (see ref.\cite{v3}).

In the QCD sum rules, the accuracy of predictions for the heavy quarkonium
masses is one order of magnitude lower than the accuracy of potential models,
$\delta m_{SR} \sim 200\div 300$ MeV. This fact is connected to that the
consideration in the QCD sum rules takes a finite number of terms in the
QCD perturbation theory for the Wilson's coefficients and a restricted set of
the quark-gluon condensates, so that the results of such noncomplete
consideration depend on an unphysical parameter, defining a scheme of
the averaging in the QCD sum rules (the number of moment for a
spectral density of current correlators or the Borel transform parameter).
An additional uncertainty is related with a modelling of a nonresonant
contribution into the current correlator, i.e. with the threshold of
hadronic continuum. Such parametric dependences lead to the low accuracy
of QCD sum rule predictions for the heavy quarkonium masses\footnote{
The QCD sum rule accuracy in calculations of the leptonic constants
($f_\psi$, $f_\Upsilon$) is higher ($\sim 20\div 25\%$), since one uses
the heavy quarkonium masses, known experimentally.}.
Moreover, the use of weight functions, defining the averaging scheme
and rapidly decreasing with the energy rise, causes a suppression of the
contribution of higher excitations in the quarkonium, so that, as a result,
this contribution is neglected, when one tries to predict the leptonic
constants of the $B_c$ meson, for example.

Recently the QCD sum rule scheme has been offered in papers of refs.
\cite{5,6}, so this scheme allows one to take into account the
spectroscopic characteristics of higher $S$-wave excitations.
Using the flavour-independence of the $S$-wave level density, one finds
the following regularities.

1) The scaling relation for the leptonic constants $f$
of $S$-wave levels of the
heavy quarkonium with the mass $M$, the reduced quark mass $\mu$ and
the fixed number of the quarkonium excitation \cite{6} is
\begin{equation}
\frac{f^2}{M}\;\biggl(\frac{M}{4\mu}\biggr)^2 = const.\;, \label{2}
\end{equation}
that is in a good agreement ($\Delta f/f \sim 5\%$) with the experimental data
\cite{9}
on the leptonic constants of $\Upsilon$, $\psi$ and even $\phi$ particles,
which are the quarkonia with the hidden flavours $(Q\bar Q)$,
so that $4\mu/M\simeq 1$, and one has \cite{5}
\begin{equation}
\frac{f^2}{M} = const.\;, \label{n2n}
\end{equation}
independently of the heavy quark flavours in the $(Q\bar Q)$ system.
Relation (\ref{n2n}) essentially differs from the scaling law for the
leptonic constants of heavy mesons $(Q\bar q)$, containing a single
heavy quark, where in Heavy Quark Effectice Theory (HQET) \cite{10}
one has
\begin{equation}
{f^2}\cdot {M} = const. \label{n3n}
\end{equation}
Law (\ref{n3n}) can be obtained from eq.(\ref{2}) in the limit
$\mu=m_q m_Q/(m_q+m_Q) \to m_q$, $m_Q \gg m_q$, $M\to m_Q$, so that
$\mu$ does not depend on the heavy quark flavour.
Eq.(\ref{2}) gives reasonable estimates for the leptonic constants of
$B$  and $D$ mesons \cite{4}, if one supposes $\mu\simeq 330$ MeV \cite{6}.

2) The scaling relation for the leptonic constants of $nS$-levels in the
quarkonium is
\begin{equation}
\frac{f^2_{n_1}}{f^2_{n_2}} = \frac{n_2}{n_1}\;, \label{3}
\end{equation}
independently of the heavy quark flavours. Eq.(\ref{3}) is in a good
agreement with the experimental data on the leptonic constants in the families
of $\psi$  and $\Upsilon$ particles \cite{9} ($\Delta f/f \le 10\%$).

3) Note, that from eq.(\ref{r11}) one can find
the relation for the mass differences of $nS$-wave levels in the
heavy quarkonium
\begin{equation}
\frac{M_n-M_1}{M_2-M_1} = \frac{\ln {n}}{\ln {2}}\;,\;\;\;n\ge 1\;,\label{m3}
\end{equation}
independently of the flavours of heavy quarks in the quarkonium. Eq.(\ref{m3})
is in a good agreement with the experimental data on the masses of
$\psi$  and $\Upsilon$ particles \cite{9} ($\delta (\Delta M)/\Delta M
\le 10\%$), too.

However, having derived relations (\ref{2}), (\ref{3}), (\ref{m3}), one has
used the phenomenological condition, stating the flavour-independence
of the heavy quarkonium level density and coming from the analysis, made
in the framework of the potential models.

In the present paper, in the framework of the offered scheme of QCD sum rules,
we derive eq.(\ref{3}) and the relation
for the $S$-wave level density for the heavy quarkonium
\begin{equation}
\frac{dM_n}{dn}(n=1) = \frac{2\bar \Lambda}{\ln{n_{th}}}\;, \label{n16}
\end{equation}
where $\bar \Lambda = m_{(Q\bar q)} - m_Q$ is the difference between the masses
of heavy meson and heavy quark, $n_{th}$ is the number of $S$-wave levels
of the $(Q\bar Q)$ heavy quarkonium below the threshold of quarkonium decay
into the heavy meson pair $(Q\bar Q) \to (Q\bar q) + (\bar Q q)$. In the
leading order, one has
\begin{equation}
\bar \Lambda = const.\;, \label{n17}
\end{equation}
with the accuracy up to power corrections over the inverse mass of heavy quark
\cite{10} (about the role of logarithmic and power corrections, see
ref.\cite{11}).

In the leading approximation, stepping from the charmonium to the bottomonium,
one can neglect a weak logarithmic variation of the number of
levels below the thershold,
\begin{equation}
\ln{n_{th}}(b\bar b) \simeq \ln{n_{th}}(c\bar c) \;. \label{n18}
\end{equation}
{}From eq.(\ref{n16})-(\ref{n18}) it follows that in the QCD sum rules one can
show that
$$
\langle T \rangle \simeq \frac{\bar \Lambda}{\ln n_{th}}\;.
$$

In Section 1 we consider the scheme of QCD sum rules with the account of
spectroscopic quantities for the heavy quarkonium and derive relations
(\ref{3}) and (\ref{n16}). In Section 2 we make the phenomenological analysis.
In the Conclusion the obtained results are summarized.

\section{Heavy Quarkonium Sum Rules}

Let us consider the two-point correlator functions of quark currents
\begin{eqnarray}
\Pi_{\mu\nu} (q^2) & = & i \int d^4x e^{iqx} \langle 0|T J_{\mu}(x)
J^{\dagger}_{\nu}(0)|0\rangle\;,
\label{1} \\
\Pi_P (q^2) & = & i \int d^4x e^{iqx}
\langle 0|T J_5(x) J^{\dagger}_5(0)|0\rangle\;,
\end{eqnarray}
where
\begin{eqnarray}
J_{\mu}(x) & = & \bar Q_1(x) \gamma_{\mu} Q_2(x)\;,\\
J_5(x) & = & \bar Q_1(x) \gamma_5 Q_2(x)\;,\\
\end{eqnarray}
$Q_i$ is the spinor field of the heavy quark with $i = c, b$.

Further, write down
\begin{equation}
\Pi_{\mu\nu}(q^2) =
\biggl(-g_{\mu\nu}+\frac{q_{\mu} q_{\nu}}{q^2}\biggr) \Pi_V(q^2)
+ \frac{q_{\mu} q_{\nu}}{q^2} \Pi_S(q^2)\;,
\end{equation}
where $\Pi_V$ and $\Pi_S$ are the vector and scalar correlator functions,
respectively. In what follows we will consider the vector and pseudoscalar
correlators: $\Pi_V(q^2)$ and $\Pi_P(q^2)$.

Define the leptonic constants $f_V$ and $f_P$
\begin{eqnarray}
\langle 0|J_{\mu}(x) |V(\lambda)\rangle & = & i \epsilon^{(\lambda)}_{\mu}\;
f_V M_V\;e^{ikx}\;,\\
\langle 0|J_{5\mu}(x)|P\rangle & = & i k_{\mu}\;f_P e^{ikx}\;,
\end{eqnarray}
where
\begin{equation}
J_{5\mu}(x)  =  \bar Q_1(x) \gamma_5 \gamma_{\mu} Q_2(x)\;,
\end{equation}
so that
\begin{equation}
\langle 0|J_{5}(x)|P\rangle  =  i\;\frac{f_P M_P^2}{m_1+m_2}\;e^{ikx}\;,
\label{9}
\end{equation}
where $|V\rangle$ and  $|P\rangle$ are the state vectors of $1^-$ and $0^-$
quarkonia, and $\lambda$ is the vector quarkonium polarization, $k$
is 4-momentum of the meson, $k_{P,V}^2 = M_{P,V}^2$.

Considering the charmonium ($\psi$, $\psi '$ ...) and bottomonium ($\Upsilon$,
$\Upsilon '$, $\Upsilon ''$ ...), one can easily show that the relation
between the width of
leptonic decay $V \to e^+ e^-$  and $f_V$ has the form
\begin{equation}
\Gamma (V \to e^+ e^-) = \frac{4 \pi}{3}\;e_i^2 \alpha_{em}^2\;
\frac{f_V^2}{M_V}\;,
\end{equation}
where $e_i$ is the electric charge of quark $i$.

In the region of narrow nonoverlapping resonances, it follows from
eqs.(\ref{1}) - (\ref{9}) that
\begin{eqnarray}
\frac{1}{\pi} \Im m \Pi_V^{(res)} (s) & = &
\sum_n f_{Vn}^2 M_{Vn}^2 \delta(s-M_{Vn}^2)\;,
\label{11} \\
\frac{1}{\pi} \Im m \Pi_P^{(res)} (s) & = &
\sum_n f_{Pn}^2 M_{Pn}^4\;\frac{1}{(m_1+m_2)^2} \delta(s-M_{Pn}^2)\;.
\end{eqnarray}
Thus, for the observed spectral function one has
\begin{equation}
\frac{1}{\pi} \Im m \Pi_{V,P}^{(had)} (s)  = \frac{1}{\pi} \Im m
\Pi_{V,P}^{(res)} (s)+ \rho_{V,P}(s, s_{th}^{V,P})\;,
\label{13}
\end{equation}
where $\rho (s,\;s_{th})$ is the continuum contribution, which is
not equal to zero at $s > s_{th}$. In the following, we will assume that the
continuum contribution is equal to the calculated perturbative part at
$s > s_{th}$, so this can lead to an additional parametric dependence
on $s_{th}$.

Moreover, the operator product expansion gives
\begin{equation}
\Pi^{(QCD)} (q^2)  = \Pi^{(pert)} (q^2)+ C_G(q^2)
\langle\frac{\alpha_S}{\pi} G^2\rangle +
C_i(q^2)\langle m_i \bar Q_i Q_i\rangle+ \dots\;,
\label{14}
\end{equation}
where the perturbative contribution $\Pi^{(pert)}(q^2)$ is labeled, and
the nonperturbative one is expressed in the form of sum
of quark-gluon condensates
with the Wilson's coefficients, which can be calculated in the QCD
perturbative theory.

In eq.(\ref{14}) we have been restricted by the contribution of vacuum
expectation values for the operators with dimension $d =4$.
For $C^{(P)}_G (q^2)$ one has, for instance, \cite{4}
\begin{equation}
C_G^{(P)} = \frac{1}{192 m_1 m_2}\;\frac{q^2}{\bar q^2}\;
\biggl(\frac{3(3v^2+1)(1-v^2)^2}
{2v^5} \ln \frac{v+1}{v-1} - \frac{9v^4+4v^2+3}{v^4}\biggr)\;, \label{15}
\end{equation}
where
\begin{equation}
\bar q^2 = q^2 - (m_1-m_2)^2\;,\;\;\;\;v^2 = 1-\frac{4m_1 m_2}{\bar q^2}\;.
\label{16}
\end{equation}
The analogous formulae for other Wilson's coefficients can be found in
ref.\cite{4}.

In the leading order of QCD perturbation theory it has been found for
the imaginary part of correlator that \cite{4}
\begin{eqnarray}
\Im m \Pi_V^{(pert)} (s) & = & \frac{\tilde s}{8 \pi s^2}
(3 \bar s s - \bar s^2 + 6m_1 m_2 s - 2 m_2^2 s) \theta(s-(m_1+m_2)^2),
\label{e38} \\
\Im m \Pi_P^{(pert)} (s) & = & \frac{3 \tilde s}{8 \pi s}
(s - (m_1-m_2)^2) \theta(s-(m_1+m_2)^2)\;,
\label{e39}
\end{eqnarray}
where $\bar s = s-m_1^2+m_2^2$, $ \tilde s^2 = \bar s^2 -4 m_2^2 s$.

The one-loop contribution into $\Im m \Pi(s)$ can be included into the
consideration (see, for example, ref.\cite{4}). However, we note that the
more essential correction is that of summing a set over the powers of
$(\alpha_S/v)$, where $v$ is defined in eq.(\ref{16})
and it is a relative quark velocity, and $\alpha_S$ is the QCD
interaction constant at the scale of characteristic average quark momentum
inside the meson \cite{v1}. In the following consideration, we will assume,
that the average square of the quark moment is
$$
\langle {\bf p}_Q^2\rangle = 2\mu \langle T\rangle\;,
$$
where $T$ is the kinetic energy and $\mu$ is the reduced mass of the system.
Since the corresponding virtualities are less than the heavy quark masses,
for the numerical estimates we
use the one-loop expression for the $\overline{MS}$ scheme
"running" coupling in QCD with three light flavours $n_f=3$
$$
\alpha_S(\langle {\bar p}_Q^2\rangle)= \frac{4\pi}{(11-2n_f/3)
\ln{\langle {\bar p}_Q^2\rangle/\Lambda^2_{QCD}}}\;,
$$
where $\Lambda_{QCD}\approx 100$ MeV reproduces $\alpha_S(m_Z^2)\simeq 0.12$
with $n_f=5$. The ${\bar p}_Q$ value is determined in accordance with the
result \cite{BLM}
$$
\langle{\bar p}_Q^2\rangle = e^{-5/3}\langle({\bf p}_Q-{\bf p'}_Q)^2\rangle
 = e^{-5/3}\;2\langle{\bf p}_Q^2\rangle\;.
$$

In ref.\cite{4}
it has been shown that account of the Coulomb-like gluonic
interaction between the quarks leads to the factor
\begin{equation}
F(v) = \frac{4 \pi}{3}\;\frac{\alpha_S}{v}\;
\frac{1}{1-\exp (-\frac{4 \pi \alpha_S}{3 v})}\;,
\end{equation}
so that the expansion of the $F(v)$ over $\alpha_S/v \ll 1$ restores,
precisely,
the one-loop $O(\frac{\alpha_S}{v})$ correction
\begin{equation}
F(v) \approx 1 + \frac{2 \pi}{3}\;\frac{\alpha_S}{v}\; \dots \label{20}
\end{equation}

The additional $\alpha_S$ correction is related with the hard gluon
contribution, it results in the factor $H$, that for the vector state
at $m_1=m_2=m_Q$ and $v\to 0$ is equal to
\begin{equation}
H_V^{m_Q} = 1 - \frac{16\alpha_S^H}{3\pi}\;,
\label{hard}
\end{equation}
where the scale of the $\alpha_S^H$ evaluation can be determined in the way,
offered in ref.\cite{BLM} (BLM). Corresponding calculations in the
$\overline{MS}$ scheme lead to the scale $e^{-11/24}m_Q\simeq  0.632 m_Q$
\cite{v1}.
The calculation of the $H$-factor for the pseudoscalar quarkonium
with $m_1\neq m_2$ was performed in ref.\cite{e1}, where one found
\begin{equation}
H_P = 1 + \frac{2\alpha_S^H}{\pi}\biggl(\frac{m_2-m_1}{m_2+m_1}
\ln{\frac{m_2}{m_1}} - 2 \biggr)\;,
\label{bra}
\end{equation}
where one takes $\alpha_S^H$ at the scale of reduced mass in the system.
Result (\ref{bra}) does not come to (\ref{hard}) at $m_1=m_2$, and, hence,
the hard gluonic corrections to the correlator of vector and pseudoscalar
currents are different. Since the nonrelativistic QCD of heavy quarks,
reformulated recently in ref.\cite{n1} and considered in ref.\cite{n2},
results in the decoupling of heavy quark spins in the leading order
interactions with gluons, the mass dependence in the $H$-factors is
determined by the corresponding renormalization in full QCD, where it comes
from the one-loop calculations of the vertex and quark self-energy
diagrams. Following ref.\cite{e1}, one straightforwardly finds
\begin{equation}
H_V = 1 + \frac{2\alpha_S^H}{\pi}\biggl(\frac{m_2-m_1}{m_2+m_1}
\ln{\frac{m_2}{m_1}} - \frac{8}{3} \biggr)\;.
\label{hardv}
\end{equation}
Results (\ref{bra}) and (\ref{hardv}) can be obtained from the
renormalization factors for vector and axial-vector currents of
the heavy quark transition $Q_1\to  Q_2$, considered in ref.\cite{vs}.
So, one must replace one heavy quark, say $Q_1$, by the antiquark $\bar Q_1$.
This leads to the subtitutions $m_1 \to -m_1$, $F_A^2 \to H_V$
and $F_V^2 \to H_P$ with the absolute value prescription for the logarith
argument.

The BLM choice of the running coupling scale for $m_1\neq m_2$ will be
considered elsewhere. Here we take the scale to be equal to
$0.63\sqrt{m_1m_2}$, that gives the estimate accuracy, suitable for the
current consideration.

In accordance with the dispersion relation one has the QCD sum rules,
which state that, in average, it is true that, at least, at $q^2 < 0$
\begin{equation}
\frac{1}{\pi}\;\int\frac{\Im m \Pi^{(had)}(s)}{s-q^2} ds = \Pi^{(QCD)}(q^2)\;,
\label{21}
\end{equation}
where the necessary subtractions are omitted. $\Im m \Pi^{(had)}(s)$ and
 $\Pi^{(QCD)}(q^2)$ are defined by eqs.(\ref{11}) - (\ref{13}) and
eqs.(\ref{14}) - (\ref{20}), respectively.
Eq.(\ref{21}) is the base to develop the sum rule approach in the forms
of the correlator function moments and of the Borel transform analysis
(see ref.\cite{4}). The truncation of the set in the right hand side of
eq.(\ref{21}) leads to the mentioned unphysical dependence of the $f_{P,V}$
values on the external parameter of the sum rule scheme.

Further, let us use the conditions, simplifying the consideration due to
the heavy quarkonium.

\subsection{Nonperturbative Contribution}

Following refs.\cite{4,s1}, we consider the $n$-th order $q^2$-derivatives
of eq.(\ref{21}) at $q^2=0$. This procedure corresponds to the calculation of
moments for the spectral densities for the current correlators. As was found
in refs.\cite{4,s1}, the ratio of $n$-th moment, calculated with the
account for the gluon condensate, essential for the heavy quarkonia,
to the $n$-th moment, calculated in the one-loop approximation of QCD
perturbative theory, is equal to
\begin{eqnarray}
\frac{A(n^{mom})}{A^{(0)}(n^{mom})} & = & 1+a(n^{mom})\alpha_S-
\nonumber \\ &&
\frac{4\pi^2}{9}\frac{n^{mom}(n^{mom}+1)(n^{mom}+2)(n^{mom}+3)}
{(2n^{mom}+5)}\frac{\langle \frac{\alpha_S}{\pi} G^2\rangle}{(2m_Q)^4}\;,
\end{eqnarray}
for the vector states at $m_1=m_2=m_Q$. The $\alpha_S$ term corresponds
to the two-loop QCD
correction. One can see, that the gluon condensate contribution will be
essential at "large" values of $n^{mom} > n_l \sim (m_Q/\Lambda)^{4/3}$,
where $\Lambda$ is the confinement (or condensate) scale. For the bottomonium
the $n_l$ value is close to 20, so that at $n^{mom} < 20$ the fraction
of the gluon condensate contribution is less than 1\%, and it rapidly
increases at $n^{mom} > 20$ (see refs.\cite{v1,v11,v2,v3}, where one can find
a manyfold discussion). Therefore, at $n^{mom} < n_l$ one can
reliably neglect the gluon contribution\footnote{
Unfortunately, the charm quark is not so heavy, and $n_l \sim 5$ \cite{4}, so
that the region of the used approximation is essentially restricted.}.

\subsection{Nonrelativistic Quark Motion}

The nonrelativistic quark motion implies that, in the resonant region, one has,
in accordance with eq.(\ref{16}),
\begin{equation}
v \to 0\;.
\end{equation}
So, one can easily find that in the leading order
\begin{equation}
\Im m \Pi_P^{(pert)}(s) \approx  \Im m \Pi_V^{(pert)}(s) \to \frac {3 v}
{8 \pi} (4\mu)^2\;,
\end{equation}
so that with account of the Coulomb factor
\begin{equation}
F(v) \simeq \frac{4 \pi}{3}\; \frac{\alpha_S}{v}\;,
\label{app}
\end{equation}
and with the hard gluon correction one obtaines
\begin{equation}
\Im m \Pi_{P,V}^{(pert)}(s) \simeq \frac{\alpha_S}{2}
(4\mu)^2\; H_{P,V}\;. \label{27}
\end{equation}

For the bottomonium the use of the $v\to 0$ limit in expression for the
Coulomb factor (\ref{app}) in the resonant region is valid with the accuracy
less than 5\%. As for the approximation of the spectral density by expression
(\ref{27}), one finds that the ratio of moments, calculated with
eqs.(\ref{e38}), (\ref{e39}) and (\ref{27}) for the resonant region
$s < s_{th}$, rapidly tends to 1, so that the maximum deviation from the unit
at moment numbers $n^{mom}\sim 1$ is close to 15\%. To decrease the value of
the error due to the nonrelativistic consideration, M.B.Voloshin
\cite{v1,v11,v3} considers $n^{mom} > 8$. In the present consideration we are
satisfied by the accuracy, when the methodics error for the leptonic
constants is close to 5\%, so that we can use nonhard restriction of
$n^{mom} > 2$, when $\Delta f/f\leq 8$\%.

Note, that in contrast to refs.\cite{v1,v11,v3}, we consider the case,
when the perturbative integrals at $s > s_{th}$ are compensated by the
hadronic continuum contributions in the sum rule equations, so that,
in practice, we consider the integration over the "resonant" region and
remember about a parametric dependence on $s_{th}$.

\subsection{Phenomenology of Hadronic Contribution}

As for the hadronic part of the correlator, one can write down for the narrow
resonance contribution
\begin{eqnarray}
\Pi_V^{(res)}(q^2) & = & \int \frac{ds}{s-q^2}\;\sum_n f^2_{Vn} M^2_{Vn}
\delta(s-M_{Vn}^2)\;,
\label{28} \\
\Pi_P^{(res)}(q^2) & = & \int \frac{ds}{s-q^2}\;\sum_n f^2_{Pn}
\frac{M^4_{Pn}}{(m_1+m_2)^2} \delta(s-M_{Pn}^2)\;,\label{29}
\end{eqnarray}
The integrals in eqs.(\ref{28})-(\ref{29}) are simply calculated, and
this procedure is generally used.

In the presented scheme, let us introduce the function of state number
$n(s)$, so that
\begin{equation}
n(M_k^2) = k\;.
\end{equation}
This definition seems to be reasonable in the resonant region.
Then one has, for example, that
\begin{equation}
\frac{1}{\pi}\; \Im m \Pi_V^{(res)}(s) = s f^2_{Vn(s)}\; \frac{d}{ds} \sum_k
\theta(s-M^2_{Vk})\;.
\end{equation}
Further, it is evident that
\begin{equation}
\frac{d}{ds} \sum_k \theta(s-M_k^2) = \frac{dn(s)}{ds}\;\frac{d}{dn} \sum_k
\theta(n-k)\;,
\end{equation}
and eq.(\ref{28}) can be rewritten as
\begin{equation}
\Pi_V^{(res)}(q^2) = \int \frac{ds}{s-q^2}\; s f^2_{Vn(s)}\;\frac{dn(s)}{ds}\;
\frac{d}{dn} \sum_k \theta(n-k)\;.
\end{equation}
Taking the average value, one finds
\begin{equation}
\Pi_V^{(res)}(q^2) = \langle\frac{d}{dn} \sum_k \theta(n-k)\rangle\;
\int \frac{ds}{s-q^2}
s f^2_{Vn(s)} \frac{dn(s)}{ds}\;.
\end{equation}
It is evident that, in average, the first derivative of step-like function
in the resonant region is equal to
\begin{equation}
\langle\frac{d}{dn} \sum_k \theta(n-k)\rangle \simeq 1\;.
\end{equation}
Thus, in the scheme, that has a rather phenomenological extent, suitable
for the large $m_Q$ limit, one has
\begin{eqnarray}
\langle\Pi_V^{(res)}(q^2)\rangle & \approx & \int \frac{ds}{s-q^2}
s f^2_{Vn(s)}\; \frac{dn(s)}{ds}\;,
\label{35} \\
\langle\Pi_P^{(res)}(q^2)\rangle & \approx & \int \frac{ds}{s-q^2}
\frac{s^2 f^2_{Pn(s)}}{(m_1+m_2)^2}\; \frac{dn(s)}{ds}\;.
\label{36}
\end{eqnarray}
To evaluate errors, related with the transformation of the sum over the
resonances to the integral over the state density, we consider the ratio
of $n^{mom}$-th moments for the vector bottomonium states and corresponding
continuous approximations for $M_n$ and $f_n$ (see figure \ref{a1}).
One can conclude that the made transformation gives the stable ratio of
moments at $n^{mom} < 20$, and this result is weakly influenced by the
variation of the continuum threshold $s_{th}$: $\delta n_{th}\simeq 0.5$, and
the quark mass threshold is more essential\footnote{Note, that this
procedure can not be considered as a way for the precision evaluation of
the heavy quark mass, since the corresponding primery accuracy in the
continuous description of the level masses is low and it can not be better
than the value of spin-spin splittings, which are neglected.}. So, this
approximation can lead to the accuracy, close to 10\% for the leptonic
constants estimation. Of course, the absolute value of the integral
representation error is related with the low numbers of excitations, and
its relative contribution rises with the growth of the moment number,
when the higher excitations become suppressed.
\setlength{\unitlength}{1mm}
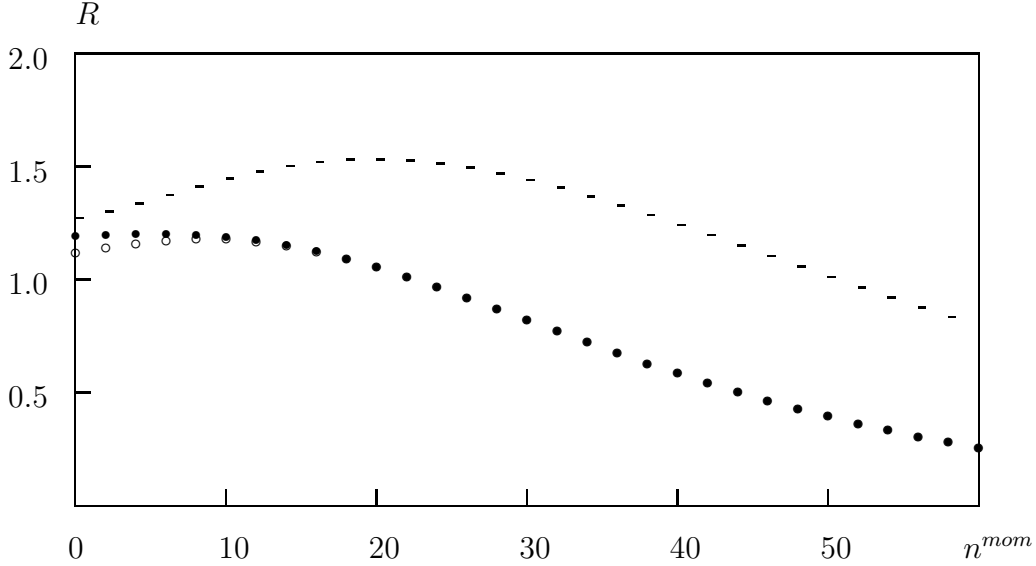
\begin{figure}[t]
\begin{picture}(140,80){}
\linethickness{0.2mm}
\put(10,10){\framebox(120,60)}
 \put( 10,45.79){\circle*{1}}
 \put( 14,45.89){\circle*{1}}
 \put( 18,46.00){\circle*{1}}
 \put( 22,46.05){\circle*{1}}
 \put( 26,45.97){\circle*{1}}
 \put( 30,45.73){\circle*{1}}
 \put( 34,45.30){\circle*{1}}
 \put( 38,44.65){\circle*{1}}
 \put( 42,43.82){\circle*{1}}
 \put( 46,42.81){\circle*{1}}
 \put( 50,41.64){\circle*{1}}
 \put( 54,40.36){\circle*{1}}
 \put( 58,38.99){\circle*{1}}
 \put( 62,37.56){\circle*{1}}
 \put( 66,36.09){\circle*{1}}
 \put( 70,34.60){\circle*{1}}
 \put( 74,33.12){\circle*{1}}
 \put( 78,31.66){\circle*{1}}
 \put( 82,30.24){\circle*{1}}
 \put( 86,28.85){\circle*{1}}
 \put( 90,27.52){\circle*{1}}
 \put( 94,26.25){\circle*{1}}
 \put( 98,25.04){\circle*{1}}
 \put(102,23.89){\circle*{1}}
 \put(106,22.81){\circle*{1}}
 \put(110,21.79){\circle*{1}}
 \put(114,20.83){\circle*{1}}
 \put(118,19.94){\circle*{1}}
 \put(122,19.11){\circle*{1}}
 \put(126,18.34){\circle*{1}}
 \put(130,17.63){\circle*{1}}
 \put( 10,43.53){\circle{1}}
 \put( 14,44.19){\circle{1}}
 \put( 18,44.76){\circle{1}}
 \put( 22,45.17){\circle{1}}
 \put( 26,45.38){\circle{1}}
 \put( 30,45.34){\circle{1}}
 \put( 34,45.04){\circle{1}}
 \put( 38,44.49){\circle{1}}
 \put( 42,43.72){\circle{1}}
 \put( 46,42.74){\circle{1}}
 \put( 50,41.61){\circle{1}}
 \put( 54,40.34){\circle{1}}
 \put( 58,38.98){\circle{1}}
 \put( 62,37.55){\circle{1}}
 \put( 66,36.08){\circle{1}}
 \put( 70,34.60){\circle{1}}
 \put( 74,33.12){\circle{1}}
 \put( 78,31.66){\circle{1}}
 \put( 82,30.24){\circle{1}}
 \put( 86,28.85){\circle{1}}
 \put( 90,27.52){\circle{1}}
 \put( 94,26.25){\circle{1}}
 \put( 98,25.04){\circle{1}}
 \put(102,23.89){\circle{1}}
 \put(106,22.81){\circle{1}}
 \put(110,21.79){\circle{1}}
 \put(114,20.83){\circle{1}}
 \put(118,19.94){\circle{1}}
 \put(122,19.11){\circle{1}}
 \put(126,18.34){\circle{1}}
 \put(130,17.63){\circle{1}}
 \put( 10, 48.20){\line(1,0){1}}
 \put( 14, 49.07){\line(1,0){1}}
 \put( 18, 50.12){\line(1,0){1}}
 \put( 22, 51.25){\line(1,0){1}}
 \put( 26, 52.40){\line(1,0){1}}
 \put( 30, 53.47){\line(1,0){1}}
 \put( 34, 54.40){\line(1,0){1}}
 \put( 38, 55.14){\line(1,0){1}}
 \put( 42, 55.66){\line(1,0){1}}
 \put( 46, 55.95){\line(1,0){1}}
 \put( 50, 56.00){\line(1,0){1}}
 \put( 54, 55.83){\line(1,0){1}}
 \put( 58, 55.44){\line(1,0){1}}
 \put( 62, 54.87){\line(1,0){1}}
 \put( 66, 54.12){\line(1,0){1}}
 \put( 70, 53.22){\line(1,0){1}}
 \put( 74, 52.20){\line(1,0){1}}
 \put( 78, 51.07){\line(1,0){1}}
 \put( 82, 49.86){\line(1,0){1}}
 \put( 86, 48.59){\line(1,0){1}}
 \put( 90, 47.26){\line(1,0){1}}
 \put( 94, 45.90){\line(1,0){1}}
 \put( 98, 44.51){\line(1,0){1}}
 \put(102, 43.12){\line(1,0){1}}
 \put(106, 41.72){\line(1,0){1}}
 \put(110, 40.34){\line(1,0){1}}
 \put(114, 38.97){\line(1,0){1}}
 \put(118, 37.62){\line(1,0){1}}
 \put(122, 36.30){\line(1,0){1}}
 \put(126, 35.01){\line(1,0){1}}
\put(30,10){\line(0,1){2}}
\put(50,10){\line(0,1){2}}
\put(70,10){\line(0,1){2}}
\put(90,10){\line(0,1){2}}
\put(110,10){\line(0,1){2}}
\put(09,3){$0$}
\put(29,3){$10$}
\put(49,3){$20$}
\put(69,3){$30$}
\put(89,3){$40$}
\put(109,3){$50$}
\put(128,3){$n^{mom}$}
\put(10,55){\line(1,0){2}}
\put(10,40){\line(1,0){2}}
\put(10,25){\line(1,0){2}}
\put(1,68){$2.0$}
\put(1,53){$1.5$}
\put(1,38){$1.0$}
\put(1,23){$0.5$}
\put(10,74){$R$}
\end{picture}
\caption{The ratio of moments for the spectral density of vector resonances
in the bottomonium, considered as the discret and continuous states,
$R=M^{dis}(n^{mom})/M^{con}(n^{mom})$. Solid and empry circles correspond to
$n_{th}=4$ and $n_{th}=4.5$, respectively, when $m_b=4.60\pm 0.01$ GeV.
Dashes show the ratio for $m_b\approx 4.64$ GeV. The fitting parameter of
the heavy quark kinetic energy equals $0.40\pm 0.03$ GeV.}
\label{a1}
\end{figure}

Note once more, that the performed transformation is purely phenomenological
representation of the experimental data.

Further, the $q^2$-derivatives of the average resonance contributions in
eqs.(\ref{35}) and (\ref{36}) can be rewritten at $q^2=0$ as
\begin{eqnarray}
\frac{(-1)^{n^{mom}}}{n^{mom}!}\; \frac{d^{n^{mom}}}{dq^{2n^{mom}}}
\langle  \Pi_V^{(res)}(0)\rangle & = &
b_V(n^{mom}) \int \frac{ds}{s^{2(n^{mom}+1)}}\;, \label{ev}\\
\frac{(-1)^{n^{mom}}}{n^{mom}!}\; \frac{d^{n^{mom}}}{dq^{2n^{mom}}}
\langle  \Pi_P^{(res)}(0)\rangle & = &
b_P(n^{mom}) \int \frac{ds}{s^{2(n^{mom}+1)}}\;, \label{ep}
\end{eqnarray}
where
\begin{eqnarray}
b_V(n^{mom}) & = & \langle s f^2_{Vn(s)} \frac{dn}{ds}\rangle|_{n^{mom}}\;,\\
b_P(n^{mom}) & = & (m_1+m_2)^{-2}\;
\langle s^2 f^2_{Pn(s)} \frac{dn}{ds}\rangle|_{n^{mom}}\;,
\end{eqnarray}
and the averaging is performed with the weight functions, depending on
the number of the spectral density moment $n^{mom}$ and shown in the right
hand sides of eqs.(\ref{ev}), (\ref{ep}). With the $n^{mom}$ growth the
$b_{V,P}$ quantities will tend to its values at the basic states
\begin{equation}
b_{V,P}  \to  \frac{1}{2} (m_1+m_2) f^2_{V_1,P_1}\;
\bigg[\frac{dM_n}{dn}(n=1)\bigg]^{-1}\;,
\label{eev}
\end{equation}
where we use $M_1\approx m_1+m_2$. Of course, the difference between the
basic state mass and the sum of quark masses becomes essential at the larger
values of the moment number $n^{mom}$, where the binding energy of quarks
is determined by the value of the gluon condensate, say, in addition to the
purely perturbative interaction.
For the bottomonium at $n^{mom}\simeq 15-20$, the accuracy of approximation
(\ref{eev}) is less than 10\%.

Next, note that in the specified region $2 < n^{mom} < n_l$, one
uses the constant value of the perturbative density, so that at the
largest admissible values of $n^{mom}$ one finds
\begin{equation}
\frac{f_{P_1,V_1}^2}{M_1} = \frac{\alpha_S}{\pi} \; \frac{dM_n}{dn}(n=1)
\; \biggl(\frac{4\mu}{M_1}\biggr)^2 H_{P,V}\;.
\end{equation}

Moreover, in the considered region of moderate numbers of the spectral
density moments, the perturbative parts can be rewritten at $q^2=0$ as
\begin{equation}
\frac{(-1)^{n^{mom}}}{n^{mom}!}\; \frac{d^{n^{mom}}}{dq^{2n^{mom}}}
\Pi_{P,V}^{(pert)}(0) = \frac{\alpha_S}{2\pi}\;16\mu^2\; H_{P,V}\;
\int \frac{ds}{s^{2(n^{mom}+1)}}\;. \label{epert}
\end{equation}
Comparing eq.(\ref{epert}) with eqs.(\ref{ev}), (\ref{ep}), one can
conclude that at $2 < n^{mom} < n_l$ the $b_{P,V}$ quantities must
be independent of $n^{mom}$ with an accuracy less than 15\%.
As the leading approximation, one can state\footnote
{Note, that the calculated imaginary part for the resonance contribution
is not assumed to be the true
physical expression, and it can not be equal to the exact hadronic
contribution. As usual in QCD sum rules, one assumes only that the
calculated expression may be, in some approximation, used for the evaluation
of the real part of the correlator. In the present paper, one believes that
the calculated part is close to the {\it average} (not exact) hadronic
one ( see eq.(\ref{r12})). In spite of the absense of explicit dependence
on the parameters of the QCD sum rule scheme ($n^{mom},\; s_{th}$), the
meaning of eq.(\ref{r12}) is strictly defined only in the specified region
of moment numbers and after the determination of scheme uncertaities,
described in the text above.}
\begin{equation}
\Im m \langle\Pi^{(hadr)}(s)\rangle = \Im m \Pi^{(QCD)}(s)\;,\label{r12}
\end{equation}
that gives with account of eqs.(\ref{27}), (\ref{35}) and
(\ref{36}) at the physical points $s_n =M_n^2$
\begin{equation}
\frac{f_{Pn,Vn}^2}{M_n} = \frac{\alpha_S}{\pi} \; \frac{dM_n}{dn}
\; \biggl(\frac{4\mu}{M_n}\biggr)^2 H_{P,V}\;. \label{38}
\end{equation}
Note, that for the real heavy quarkonia $(\bar b b)$, $(\bar b c)$ and
$(\bar c c)$, the scale-dependent part in the right hand side of eq.(\ref{38})
is approximately constant
$$
\alpha_S H_{P,V}\approx const.\;,
$$
and it is practically independent of the total spin of quarks,
so that
\begin{equation}
f_{Vn} \simeq f_{Pn} = f_n\;,\label{40}
\end{equation}
with the accuracy $\Delta f/f \leq 5$\%.
Thus, one can conclude that for the heavy quarkonia the QCD sum rule
approximation gives
the identity of $f_P$ and $f_V$ values for the pseudoscalar
and vector states.

Eq.(\ref{38}) differs from the ordinary sum rule scheme because it does not
explicitly contain the parameters, which are external to QCD. The quantity
$dM_n/dn$ is purely phenomenological. It defines the average mass difference
between the nearest levels with the identical quantum numbers.

Further, as it has been shown in ref.\cite{11a}, in the region of average
distances between the heavy quarks in the charmonium and the bottomonium,
\begin{equation}
0.1\; fm < r < 1\;fm\;, \label{2.1}
\end{equation}
the QCD-motivated potentials allow the approximation in the form of
logarithmic law \cite{2} with the simple scaling properties, so
\begin{equation}
\frac{dn}{dM_n} = \phi(n)\;,\label{2.2}
\end{equation}
i.e. the density of heavy quarkonium states with the given quantum
numbers do not depend on the heavy quark flavours.

In ref.\cite{5} it has been found, that relation (\ref{2.2}) is valid
with the accuracy up to small logarithmic corrections over the
reduced mass of quarkonium, if one makes the quantization of
$S$-wave states for the quarkonium with the Martin potential by the
Bohr-Sommerfeld procedure.

Thus, as it has been shown in refs.\cite{5,6},
for the leptonic constants of $S$-wave quarkonia, the scaling relation
takes place
\begin{equation}
\frac{f_n^2}{M_n}\; \biggl(\frac{M_n}{4\mu}\biggr)^2 = c_n\;, \label{2.3}
\end{equation}
independently of the heavy quark flavours.

This conclusion can be drawn with no use of the potential models, since
the approximation, when the difference between the level masses in the
heavy quarkonia is flavour-independent, is phenomenological observation,
leading to the flavour-independence of the state density.

Further, the phenomenology for the quarkonium masses in the framework of the
potential models leads to the simple scaling relation for the state density
(\ref{r11}).

Note, that from eq.(\ref{r11}) one finds
\begin{equation}
\frac{M_n-M_1}{M_2-M_1} = \frac{\ln {n}}{\ln {2}}\;,\label{2.9}
\end{equation}
and
\begin{equation}
{M_2-M_1} = \frac{dM_n}{dn}(n=1)\; {\ln {2}}\;.\label{2.10}
\end{equation}

Eq.(\ref{2.9}) for the differences of $nS$-wave level masses of the heavy
quarkonium does not contain external parameters and it allows direct
comparison with the experimental data on the masses of particles in
the $\psi$  and $\Upsilon$ families \cite{9}.

\begin{figure}[t]
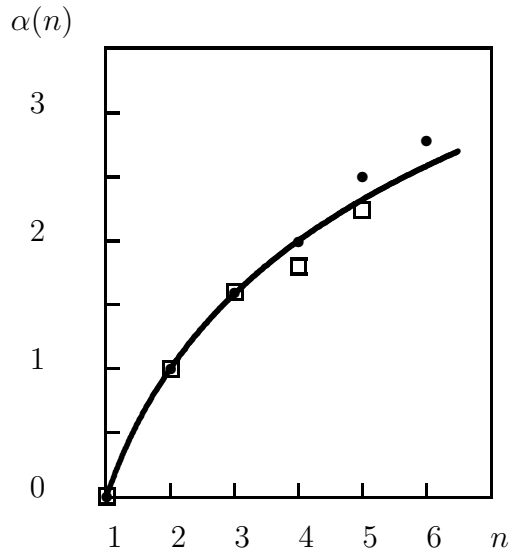

\begin{center}

\end{center}
\caption{The experimental values of $nS$ bottomonium (solid dots)
and charmonium (empty boxes) mass differences
$\alpha(n) =(M_n-M_1)/(M_2-M_1)$
and the dependence in the present model $\alpha(n)=\ln{n}/\ln{2}$.}
\label{fm}
\end{figure}

Dependence (\ref{2.9}) and the experimental values for the relations of
heavy quarkonium masses are presented on figure \ref{fm},
where one neglects the spin-spin splittings.

Note, the $\psi(3770)$ and $\psi(4040)$ charmonium states suppose to be the
results of the $3D$- and $3S$-states mixing, so that the $D$-wave dominates
in the $\psi(3770)$ state, and the mixing of the $3D$- and $3S$-wave functions
is accompanied by a small shifts of the masses, so that we have
supposed $M_3= M_{\psi(4040)}$ and $f_3^2=f^2(3770)+f^2(4040)$.

As one can see from the figure \ref{fm}, relation (\ref{2.9}) is in a
good agreement with the experimental data.
\setlength{\unitlength}{0.85mm}\thicklines

This relation in the hadronic spectra must result in the specific relarions
for the leptonic constants, if one considers the QCD sum rules.

Using eqs.(\ref{r11}) and (\ref{38}), one gets
\begin{equation}
\frac{f^2_{n_1}}{f^2_{n_2}} = \frac{n_2}{n_1}\;. \label{2.6}
\end{equation}

First, note that eq.(\ref{2.3}), relating the leptonic constants of
different quarkonia, turns out to be certainly valid
for the quarkonia with the hidden
flavour ($c\bar c$, $b\bar b$), where $4\mu/M\simeq 1$ \cite{5,6} (see
Table \ref{tm1}).

\begin{table}[t]
\caption{The experimental values of leptonic constants (in MeV)
for the quarkonia in comparison with the estimates of present model.}
\label{tm1}
\begin{center}
\begin{tabular}{||l|c|c||}
\hline
quantity & exp. & present \\
\hline
$f_\phi$ & $232\pm5$ & $235\pm25$ \\
$f_\psi$ & $409\pm13$ & $408\pm40$ \\
$f_\Upsilon$ & $714\pm14$ & $714\pm45$ \\
\hline
\end{tabular}
\end{center}
\end{table}

Second, eq.(\ref{2.3}) gives estimates of the leptonic constants for
the heavy $B$ and $D$ mesons, so these estimates are in a good agreement with
the values, obtained in the framework of other schemes of the QCD
sum rules \cite{4}.

\begin{figure}[t]
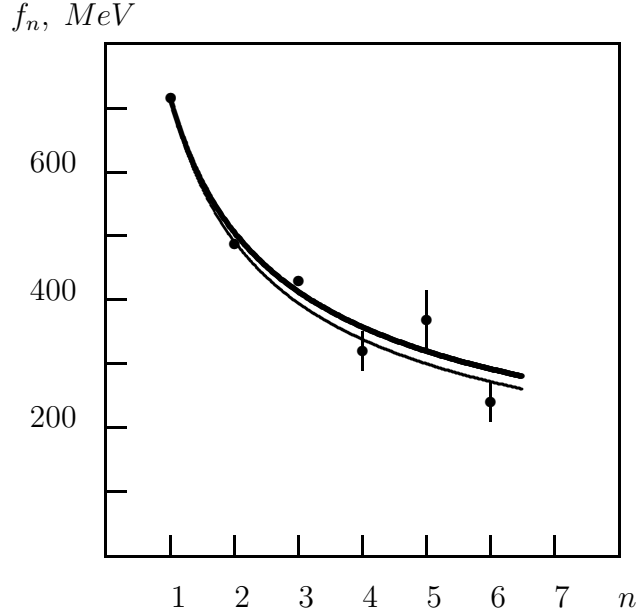

\begin{center}

\end{center}
\caption{The calculated dependence of $nS$ bottomonium leptonic
constants and the experimental values of $f_{\Upsilon(nS)}$.}
\label{fh1}
\end{figure}

\begin{figure}[t]
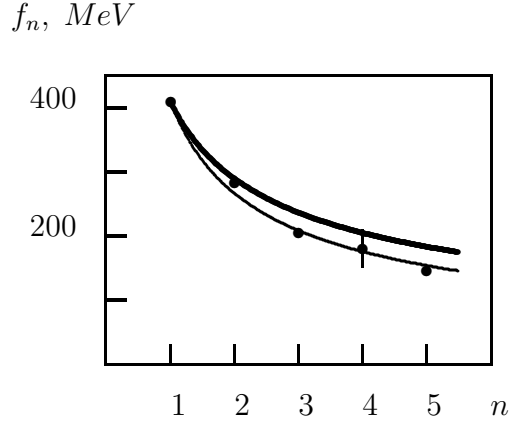

\begin{center}

\end{center}
\caption{The calculated dependence of $nS$ charmonium leptonic
constants and the experimental values of $f_{\psi(nS)}$.}
\label{fh2}
\end{figure}

Third, taking a value of the $1S$-level leptonic constant as the input one, we
have calculated the leptonic constants of higher $nS$-excitations in the
charmonium and the bottomonium and found a good agreement with the experimental
values (see the thick line curvers on figures \ref{fh1}, \ref{fh2}),
where we do not take into account
the deviation of $4\mu/M$ quantity from the unit, however this effect can be
essential for the $(\bar c c)$ system. One can easily find the modification
form of eq.(\ref{2.6})
\begin{equation}
\frac{f^2_{n_1}}{f^2_{n_2}} = \frac{n_2}{n_1}\;\frac{M^2_{n_2}}{M^2_{n_1}}\;.
\label{2.6'}
\end{equation}
The leptonic constant values, rescaled under relation (\ref{2.6'}), are
presented as the thin line curvers on figures \ref{fh1}, \ref{fh2}, so that
one can see the uncertainty, caused by the small excitation approximation.

\setlength{\unitlength}{0.85mm}\thicklines

These facts show that the offered scheme can be quite reliably applied
to the systems with the heavy quarks.

Further, using eq.(\ref{2.6}), at $q^2=0$ one can write down
\begin{equation}
\int_{s_i}^{s_{th}} \frac{ds}{s}\; s\; f^2_{n(s)}\; \frac{dn}{ds}=
f^2_1 \int_{s_i}^{s_{th}} ds\; \frac{d\ln{n}}{ds} = f^2_1\; \ln{n_{th}}\;.
\label{2.11}
\end{equation}
On the other hand, in the leading approximation, one gets
\begin{equation}
\int_{s_i}^{s_{th}} ds\; \frac{\alpha_S}{2\pi}\;
\biggl(\frac{4\mu}{M}\biggr)^2 H_V=
\frac{\alpha_S}{2\pi}\; \biggl(\frac{4\mu}{M}\biggr)^2 H_V\; (s_{th}-s_i)\;,
\label{2.12}
\end{equation}
and, further,
\begin{equation}
s_{th}-s_i \simeq 2M \Delta E\;, \label{2.13}
\end{equation}
where $\Delta E = E_{th} - m_Q - m_{Q'}$ is the difference between the
threshold energies for the decay $(Q\bar Q') \to (Q\bar q) + (\bar Q' q)$ and
the $(Q\bar Q')$ pair production.

In HQET \cite{10} one has
\begin{equation}
\Delta E = 2\bar \Lambda + O(1/m_Q)\;, \label{2.14}
\end{equation}
i.e. in the leading approximation one can take $\Delta E \simeq 2\bar \Lambda$,
being a constant value, independent of the heavy quark flavours.

Then one finds
\begin{equation}
\frac{f_1^2}{M} = \frac{\alpha_S}{\pi}
\; \biggl(\frac{4\mu}{M}\biggr)^2 H_V \; \frac{2\bar \Lambda}{\ln{n_{th}}}\;.
\label{2.15}
\end{equation}
Comparing eq.(\ref{38}) and eq.(\ref{2.15}), one can easily find that
in the leading approximation
\begin{equation}
\frac{dM_n}{dn}(n=1) = \frac{2\bar \Lambda}{\ln{n_{th}}}\;. \label{2.16}
\end{equation}
This means, that the average kinetic energy of heavy quarks equals
$$
\langle T\rangle \simeq \frac{\bar \Lambda}{\ln{n_{th}}}\;.
$$
Having derived eq.(\ref{2.16}), one has assumed, that

1) the binding energy of quarks in the $1S$-state is negligibly small, than
the excitation energy of $nS$-levels
\begin{equation}
E_1 \ll \bar \Lambda \sim \frac{dM}{dn}\;, \label{2.17}
\end{equation}

2) the excitation energy of levels is small in comparison with the quark masses
\begin{equation}
\bar \Lambda \sim \frac{dM}{dn} \ll m_Q\;, \label{2.18}
\end{equation}
so that $\sqrt{s} \sim M$,

3) in the leading approximation the hadronic continuum threshold is
determined by the masses of heavy mesons
\begin{equation}
\sqrt{s_{th}} \simeq M_{(Q\bar q)} + M_{(\bar Q' q)} \simeq
m_Q +m_{Q'} + 2 \bar \Lambda \;, \label{2.19a}
\end{equation}

4) the number of states below the threshold is finite and weakly depends on
the heavy quark flavours.

Thus, from eq.(\ref{2.16}) one can conclude that in the
framework of the QCD sum rules, one gets the universal regularity for the
density of $S$-wave quarkonium levels, so, in the leading approximation,
the relation does not depend on the heavy quark flavours
\begin{equation}
\frac{dM_n}{dn}(n=1) = const. \label{2.19}
\end{equation}

\section{Numerical Analysis}

Relation (\ref{2.16}) is got in the leading approxomation over the
inverse mass of heavy quarks, when one can neglect the spin-dependent
splittings. Therefore, for the $\bar \Lambda$ estimate we will use the values
of $S$-level masses of the quarkonia $(\bar c c)$ and $(\bar b b)$.

One can easily show

\begin{eqnarray}
m(n^3S_1) & = & m(nS) + \frac{1}{4} \Delta m(nS)\;,\label{2.20}\\
m(n^1S_0) & = & m(nS) - \frac{3}{4} \Delta m(nS)\;,\nonumber
\end{eqnarray}
where $\Delta m(nS)$ is proportional to the leptonic constant squared
$f_{nS}^2$ \cite{1,2,3,11a}, so that from the previous Section it follows
\begin{equation}
\Delta m(nS) = \frac{\Delta m(1S)}{n}\;. \label{2.21}
\end{equation}
{}From the experimental data  one has $\Delta m_\psi(1S) =117$ MeV, and,
taking into account eqs.(\ref{2.20}, \ref{2.21}), one gets
\begin{eqnarray}
m_\psi(1S) & = & 3.068\;\;GeV\;,\nonumber\\
m_\psi(2S) & = & 3.670\;\;GeV\;,\nonumber\\
m_\Upsilon(1S) & = & 9.440\;\;GeV\;,\label{2.22}\\
m_\Upsilon(2S) & = & 10.012\;\;GeV\;,\nonumber
\end{eqnarray}
where we have taken into account that
$$
\Delta m_\Upsilon = \Delta m_\psi\; \frac{\alpha_S(\Upsilon)}
{\alpha_S(\psi)}\;,
$$
and $\alpha_S(\Upsilon)/\alpha_S(\psi)\simeq 3/4$ \cite{12}.
{}From eq.(\ref{2.22}) one has
\begin{eqnarray}
(M_2-M_1)|_\psi & \simeq & 0.602\;\;GeV\;,\label{2.23}\\
(M_2-M_1)|_\Upsilon & \simeq & 0.572\;\;GeV\;,\nonumber
\end{eqnarray}
i.e. in average one has
\begin{equation}
\langle M_2-M_1\rangle \simeq  0.587\pm 0.015\;\;GeV\;.\label{2.24}
\end{equation}

In the $\Upsilon$ family, where the leading approximation over the
inverse heavy quark mass must be the most reliable, one has
\begin{equation}
n_{th} = 4\;.\label{2.25}
\end{equation}
Then
\begin{equation}
\bar \Lambda  = \langle M_2-M_1\rangle
\label{2.26}
\end{equation}
Accounting for the variation of $n_{th}$ one gets
\begin{equation}
\bar \Lambda  = 0.59\pm 0.07\;\;GeV\;.
\label{2.27a}
\end{equation}
Note, that the threshold energy of the hadronic continuum can be greater
than the double mass of heavy meson, since a production of light hadrons
(such as $\pi$ mesons) can be essential. In the last case the level
spacing is rewritten as
\begin{eqnarray}
M_2 -M_1 & = & \frac{2(\bar \Lambda +\delta E_{th})\ln 2}
{\ln(n_{th}+\delta n_{th})} \nonumber\\
& \approx & \frac{2\bar \Lambda\ln 2}{\ln n_{th}}\;
(1+\frac{\delta E_{th}}{\bar \Lambda})+\frac{2\bar \Lambda\ln 2}{n_{th}}
\delta n_{th}\;,
\end{eqnarray}
so that the variations of $E_{th}$ and $n_{th}$ correlate with each other.
The corresponding uncertainty of  $\bar \Lambda$ due to the variation
of the hadronic continuum threshold is included in the $\bar \Lambda$
estimation.

Estimate (\ref{2.27a}) of the important parameter in HQET is in a good
agreement with the estimates, made in the QCD sum rules for the heavy mesons
\cite{11}
\begin{equation}
\bar \Lambda  \simeq  0.57\pm 0.07\;\;GeV\;.\label{2.27}
\end{equation}
The accuracy of estimate
(\ref{2.27a}) is within the limits of
the accuracy $\delta \bar \Lambda \sim 20$ MeV, that can be achieved, because
of the nonperturbative corrections in QCD \cite{11}.

{}From eq.(\ref{2.27a}) one finds the estimate
\begin{equation}
\frac{dM}{dn}(n=1) \simeq  0.85\pm 0.09\;\;GeV\;,\label{2.28}
\end{equation}
that is slightly greater than the estimates, made in papers of ref.\cite{13},
where $dM/dn \simeq 0.75$ GeV was determined in the polinomial interpolation
of heavy quarkonium masses.

Finally, one has
$$
\langle T\rangle = 0.43\pm 0.04\;\;GeV\;.
$$

Using the $1S$-level masses of $B$ and $D$ mesons, one can write down the
expression for the HQET pole mass of heavy quarks
$$
m_Q = m_{Q\bar q}(1S) - \bar \Lambda -\frac{\mu_\pi^2}{2m_{Q\bar q}(1S)}\;,
$$
where $\mu_{\pi}^2 = 0.4\pm 0.1$ GeV$^2$ \cite{mu} is the average square of
the heavy quark momentum inside the heavy meson. Note, that the
$O(1/m_Q)$ correction to the heavy quark mass can be valuable for the charm
quark. So, one finds
\begin{eqnarray}
m_b & = & 4.69 \pm 0.08\;\; GeV\;,\\
m_c & = & 1.30 \pm 0.12\;\; GeV\;.
\end{eqnarray}
However, the mass-dependent correction to the HQET pole mass of the charm
quark is only $1.5 \sigma$ deviation from the leading approximation $O(1)$
over the inverse heavy quark mass, so that the central value of $m_c$ in the
equation above can be reasonably taken as $m_c=1.40\pm 0.12$, that is more
suitable for the considered approximation of a low binding in the quarkonium
and it agrees with the value, obtained in the consideration of the integral
representation for the resonance contribution, described above. This effect
leads also to poor accuracy in the extraction of $\bar \Lambda$ value
from the charmonium spectroscopy, where the region of $n^{mom}$, suitable for
the made approximations, is strongly restricted.

Next, using the scaling relation, one can get the prediction
\begin{equation}
f_{B_c} = 385\pm 25\;\;MeV\;,
\end{equation}
that is in agreement with other estimates, performed in QCD sum rules \cite{f}.

The data on the $\psi$ and $\Upsilon$ particles \cite{9} give
\begin{equation}
a_Q = \alpha_S\;\biggl(1 - \frac{16\alpha_S^H}{3\pi}\biggr)\;
\biggl(\frac{2m_Q}{M_1}\biggr)^2 = 0.21\pm 0.01\;,
\end{equation}
that must be compared with the estimates, corresponding to the
described choice of scale parameters
\begin{eqnarray}
a_b & = & 0.21\pm 0.01\;,      \\
a_c & = & 0.19\pm 0.04\;,
\end{eqnarray}
where the uncertainty in the result for the charm quark is basically
related with the deviation of $(2m_c/M_1)^2$ factor from the unit as well as
with the large contribution of $O(1/m_c)$ term in the value of $c$-quark mass.
So, the offered relations are in a good agreement with the current data in the
limits of the approach accuracy.

\section*{Conclusion}

In the framework of the QCD sum rules and with the use of the phenomenological
quasiclassical relation for the state density, the expression for the density
of $S$-wave levels of the heavy quarkonium is derived
$$
\frac{dM_n}{dn}(n=1) = \frac{2\bar \Lambda}{\ln{n_{th}}}\;,
$$
that in the leading approximation does not depend on the heavy quark flavours.
Here, $\bar \Lambda = m_{(Q\bar q)}-m_Q$ and $n_{th}$ is the number of
$nS$-levels below the threshold of $(Q\bar Q') \to (Q\bar q)+(\bar Q' q)$
decay. The analysis of the spectroscopic data on the charmonium and
bottomonium allows one to do the estimate
$$
\bar \Lambda \simeq  0.59\pm 0.07\;\;GeV\;,
$$
that is in a good agreement with the recent estimates from the QCD sum
rules for the heavy mesons.

The phenomenology of quarkonium spectra in the framework of the
potential models leads to the simple scaling expression for the state
density, that in the QCD sum rules results in the scaling relation for
the leptonic constants of excited $nS$-states
$$
\frac{f^2_{n_1}}{f^2_{n_2}} = \frac{n_2}{n_1}\;.
$$

The author expresses the gratitudes to G.P.Pron'ko for exciting discussions
and to N.Paver and M.B.Voloshin for fruitful controversies.

This work is partially supported by the ISF grant NJQ000 and the programm
"Russian State Stipendia for Young Scientists".

\vspace*{0.4cm}
\hfill Recieved April 17, 1995
\end{document}